\documentclass[10pt]{article}
\usepackage[utf8]{inputenc}

\title {An Improved Inertia Principle}
\author{Rodrigo Medina \\
\small Centro de F\'{\i}sica\\[-0.8ex]
Instituto Venezolano de
Investigaciones Cient\'{\i}ficas\\
\small Apartado 20632 Caracas 1020-A, Venezuela\\
\small \texttt{rmedina@ivic.gob.ve}\\
\and   J. Stephany\\
\small  Departamento de F\'{\i}sica\\[-0.8ex]
\small Universidad Sim\'{o}n Bol\'{\i}var\\[-0.8ex]
\small Apartado 89000, Caracas 1080A, Venezuela.\\
\small \texttt{stephany@usb.ve}
}

\date{}
\begin{document}

\maketitle
\begin{abstract}
We show  that for isolated relativistic systems with spin  the conservation of total angular momentum  implies  that, instead of the center of mass, it is a modified center of mass and spin which behaves inertially. This requires a change in the the statement of the Principle of Inertia. 
\end{abstract}

We show in this letter how conservation of total angular momentum in systems with spin implies a modification of the statement of the Principle of Inertia. The Principle of Inertia, Newton’s first law is a cornerstone  of Classical and Relativistic Mechanics, where it appears entangled with the  definition and postulated  existence of inertial frames of reference. The standard statement of the principle is: “There are inertial frames of reference in which free particles move with constant velocity”. The concept of free particle is needed for its formulation, understood as a body whose dimensions are negligible with respect to all other relevant distances, and which does not interact with other particles or fields. Any body  chosen to play this role may be explored to a scale in which it appears composite. This is true even for electrons which at tiny scales are affected by vacuum polarization. To overcome the limitation associated to the particle 
concept, the postulate is often rephrased in the apparently more general form which states that: “There are inertial frames of reference in which the center of mass of isolate systems move  with constant velocity”. Then objects much bigger than electrons like hadrons, nuclei, atoms or even rockets may be considered when exploring the inertiality of a reference frame. But this statement is in fact a theorem which depends  on other aspects of the theoretical description of the system. In Classical Mechanics an isolated system is one for which the total external force and torque vanish. The center of mass of an isolated system  moves  with constant velocity if the sum of all the forces that any of its parts exerts on all others is zero. This condition is provided automatically when  Newton's Third Law, the Action and Reaction Principle, holds. 

In Relativistic Physics an isolated system has a divergence less energy-momentum tensor $T^{\mu\nu}$. The four-momentum vector $P^{\mu}=\frac{1}{c}\int T^{\mu 0} dv$ is conserved: $\frac{d}{dt} P^{\mu}=0$. In particular the energy $U=cP^0$ is conserved. The center of mass defined by $X^i_T=\frac{1}{U}\int x^i T^{00}dv$ moves with constant velocity when the orbital  angular momentum $L^{\mu\nu}=\frac{1}{c}\int (x^\mu T^{\nu0}-x^\nu T^{\mu0})dv$ is conserved:$\frac{d}{dt} L^{\mu\nu}=0$. In this case  one has,
\begin{equation}
\frac{d}{dt} L^{0i}=\frac{1}{c} \frac{d}{dt}\int(x^{0}T^{i0}-x^i T^{00})dv=\frac{U}{c}(\frac{c^2}{U}P^i- \frac{d}{dt}X^i_T)=0\ .
\end{equation}
The velocity of the  center of mass is $\frac{c^2}{U}P^i$. This theorem is called the center of mass motion theorem (CMMT). When $T^{\mu\nu}$ is symmetric, the orbital angular momentum is conserved and CMMT holds.

Through the years there were hints that  in systems with electromagnetic interaction, the CMMT may not hold \cite{Cullwick1952,Balazs1953,Taylor1965,SJ1967}. Regretfully they remained unnoticed in the  bulk of the Abraham-Minkowski controversy about the momentum density of the electromagnetic field, or were dismissed in favor of more speculative approaches like modifications of Lorentz force or the hidden momentum hypothesis \cite{SJ1967}. This was motivated probably by  the aversion which generates in many people the possibility that the Inertia Principle could be affected. In a recent work \cite{MedandSa} we point out that  the CMMT  is indeed not  valid when the energy momentum tensor is not symmetric and discuss an explicit example of an isolated system were the dynamical equations imply that it is violated. The example is very simple. It consists of a charge in the center of a toroidal magnet with circular magnetization. When the magnet is heated above Curie point it is straightforward to show, using 
Maxwell equations and the Lorentz force, that the induced electric field accelerates the charge without any reaction on the magnet. Consequently the center of mass 
of the system also accelerates. Since as mentioned, this has implications on the Inertia Principle in this communication we want to discuss from a more general point of view the reason  that allow the violation of the CMMT and how an improved inertia principle emerges. 

The CMMT is not always valid for the simple reason that orbital angular momentum is not always conserved. As has been know for more than eighty years what is conserved in general is the total angular momentum which includes spin. Fortunately an improved theorem may be formulated. Let be $S^{\mu\nu}$ the spin tensor of the system under consideration and suppose that the total angular momentum $J^{\mu\nu}=L^{\mu\nu}+S^{\mu\nu}$ is conserved:$\frac{d}{dt} J^{\mu\nu}=0$. Define $X^i_S=-\frac{c}{U}S^{0i}$. By the conservation of the total angular momentum one has,
\begin{eqnarray}
\label{CMSMT}
\frac{d}{dt}X^i_S&=&\frac{c}{U} \frac{d}{dt} L^{0i}= \frac{1}{U}\frac{d}{dt}\int \big[x^0 T^{i0}-x^iT^{00}\big]dv\nonumber\\
&=&\frac{c^2 P^i}{U}-\frac{d}{dt}X^i_T\ .
\end{eqnarray}
This shows that it is the center of mass and spin defined by $X^i_\Theta=X^i_T+X^i_S$ which moves with constant velocity. Consequently, to open room to systems with non vanishing spin which include besides magnets all the particles of the standard model but the Higgs, the inertia principle should be enunciated in the form: “There are inertial frames of reference in which the center of mass and spin of isolate systems move with constant velocity”. This slight modification  provides the accurate statement of the inertia principle for systems with spin. Since $S^{0i}$ has went unnamed for many years we propose that it should be called the spin arrow.  $L^{0i}$ would then be the orbital angular momentum arrow. 

Although our discussion is classical, our analysis may be applied to particles as it is done with the usual CMMT. Consider an electron moving with rapidity $\beta=\frac{v}{c}$ in the $x^1$ direction of the laboratory frame. Suppose that in its rest frame the spin of the electron is polarized in the direction of $x^3$, i.e $S'^{12}=-S'^{21}=\frac{\hbar}{2}$. Since in this case $S^{\mu\nu}$ is  independently conserved, using the Lorentz transformation in the laboratory frame the non vanishing components of the spin tensor are $S^{12}=-S^{21}=\frac{\gamma\hbar}{2}$ and $S^{02}=-S^{20}=\frac{\gamma\beta\hbar}{2}$ with $\gamma=(1-\beta^2)^{-1/2}$. The only contribution of the spin to the center of mas and spin is $X^2_S=-\frac{\beta\hbar}{2m_ec}$. The center of mass and spin moves in this case at a distance $\frac{\beta\lambda_e}{4\pi}$ parallel to the trajectory of the particle with $\lambda_e$ the Compton wavelenght of the electron. Even for $\beta\rightarrow 1$ this is a small distance and the kinematic 
consequences of the difference between the center of mass and  spin and the center of mass may be difficult to detect. For an electron neutrino the effect is at least 250000 times larger.

In the example of Ref.\cite{MedandSa} spin is present in the magnetization of matter but also in the electromagnetic field. The spin dynamics of that system is certainly complicated and involves both the changing magnetization and the spin density of the field but the general result Eq.(\ref{CMSMT}) is independent of the  details. The force that accelerates the magnet in this example is very small and also difficult to detect, but not necessarily out of reach of experiment.

It should be noted that the possibility of orbital angular momentum to be converted in spin and vice versa is determined by the coupling of spin with the non-symmetric part of the energy-momentum tensor. In Ref. \cite{MedandSb} we found from first principles that the true energy-momentum tensor of the electromagnetic field in matter is not necessarily symmetric. Our result  confirms Minkowski's expression $\mathbf{g}_{\mathrm{Min}} = c^{-1} T^{i0}_F \hat{\mathbf{e}}_i =\frac{1}{4\pi c}\mathbf{D}\times\mathbf{B}$ as the momentum density of the field. This in particular determines that the momentum of photons with energy $E$ in matter is  $nE/c$ with $n$ the refraction index as most of the experiments suggest \cite{Campbell,Leonhart2006}.

Magnetic moment in the rest frame is proportional to spin. In a moving system it manifests itself also  as  electric dipole moment which should then be  proportional to spin arrow. Unlike leptons, molecules as well as macroscopic samples of polarized dielectrics may have an electric dipole-moment in their rest-frame. An intriguing question is whether electric dipole moment in the rest frame of a system may always be related with a non-vanishing spin or orbital angular momentum arrow. If not, a second question is what is the physical meaning  of the magnetization-like spatial companion of the dipolar polarization vector which appears in a moving frame. If the answer to the first question is on the positive the center of mass and spin of a polarized dielectric sample would be displaced from its geometrical center.

Finally we should mention that $X^i_\Theta$ corresponds to the usual definition of the center of mass, if one uses the Belinfante-Rosenfeld tensor which is a symmetric combination of the energy-momentum tensor and the spin density often confused with the energy momentum tensor. This confusion occurs mainly because it appears as the source of Einstein equations in the flat limit when field theories are coupled to the gravitational field. We think that the correct interpretation of this fact is that spin is also a source of the gravitational field in general relativity and not that the true energy-momentum tensor is Belinfante-Rosenfeld tensor. This interpretation has interesting consequences in cosmological models where spin may give a contribution to dark matter and in models of magnetized compact objects.  The relation  of the inertia principle and Belinfante-Rosenfeld tensor will be discussed in a forthcoming paper \cite{MedandSe}.

\end{document}